\documentclass[fleqn,10pt]{SelfArx}
\usepackage[english]{babel}

\usepackage{listings}
\usepackage{xcolor}

\definecolor{codegreen}{rgb}{0,0.6,0}
\definecolor{codegray}{rgb}{0.5,0.5,0.5}
\definecolor{codepurple}{rgb}{0.58,0,0.82}
\definecolor{backcolour}{rgb}{0.95,0.95,0.92}

\lstdefinestyle{mystyle}{
    backgroundcolor=\color{backcolour},   
    commentstyle=\color{codegreen},
    keywordstyle=\color{magenta},
    numberstyle=\tiny\color{codegray},
    stringstyle=\color{codepurple},
    basicstyle=\ttfamily\footnotesize,
    breakatwhitespace=false,         
    breaklines=true,                 
    captionpos=b,                    
    keepspaces=true,                 
    numbers=left,                    
    numbersep=5pt,                  
    showspaces=false,                
    showstringspaces=false,
    showtabs=false,
    tabsize=2
}
\usepackage{pgfplots}
\pgfplotsset{width=8cm, compat=1.17}

\usepackage[style=numeric,sorting=none,backend=biber]{biblatex}
\usepackage{csquotes}

\usepackage{enumitem}

\definecolor{color1}{RGB}{0,0,90} 
\definecolor{color2}{RGB}{0,20,20} 

\usepackage{url}

\usepackage{hyperref}
\hypersetup{
	hidelinks,
	colorlinks,
	breaklinks=true,
	urlcolor=color2,
	citecolor=color1,
	linkcolor=color1,
	bookmarksopen=false,
	pdftitle={LLM Interactive Optimization of Open Source Python Libraries - Case Studies and Generalization},
	pdfauthor={Florath, Andreas},
}

\JournalInfo{V2}
\Archive{\ }

\PaperTitle{LLM Interactive Optimization of Open Source Python Libraries - Case Studies and Generalization}

\Authors{Andreas Florath\textsuperscript{1}*}
\affiliation{\textsuperscript{1}\textit{Group Technology, Deutsche Telekom AG, Bonn, Germany}}
\affiliation{*\textbf{Corresponding author}: Andreas.Florath@telekom.de}

\newcommand{\keywordname}{Keywords}

\Abstract{With the advent of large language models (LLMs) like GPT-3,
    a natural question is the extent to which these models can be
    utilized for source code optimization.
    This paper presents methodologically stringent case studies applied
    to well-known open source python libraries pillow and numpy.

    We find that contemporary LLM ChatGPT-4 (state September and October 2023)
    is surprisingly adept at optimizing energy and compute efficiency.
    However, this is only the case in interactive use,
    with a human expert in the loop.

    Aware of experimenter bias, we document our qualitative approach in detail,
    and provide transcript and source code. We start by providing a detailed
    description of our approach in conversing with the LLM to optimize
    the \texttt{\_getextrema} function in the pillow library, and a quantitative
    evaluation of the performance improvement.

    To demonstrate qualitative replicability, we report further attempts
    on another locus in the pillow library, and one code locus
    in the numpy library, to demonstrate generalization within and beyond a library.

    In all attempts, the performance improvement is significant (factor up to 38).
    We have also not omitted reporting of failed attempts (there were none).

    We conclude that LLMs are a promising tool for code optimization in open source
    libraries, but that the human expert in the loop is essential for success.
    Nonetheless, we were surprised by how few iterations were required to achieve
    substantial performance improvements that were
    not obvious to the expert in the loop.

    We would like bring attention to the qualitative nature of this study,
    more robust quantitative studies would need to introduce a layer of selecting
    experts in a representative sample --- we invite the community to collaborate.}

\begin{document}

\maketitle 

\tableofcontents 

\thispagestyle{empty} 


\section{Introduction}

\subsection{Aims}
With progress in large language models (LLMs) rapidly accelerating,
much attention has been devoted to their potential in programming,
autonomously or in collaboration with humans.

Significant potential could be expected in optimizing energy or compute efficiency
of source code, with a positive impact on the environment, code health, and cost.

While a plethora of research focuses on development of LLMs and algorithms for
coding tasks, the area of code optimization has received comparably little attention
(see \ref{subsec:PriorArt}~\nameref{subsec:PriorArt}).

It is even more surprising that interactive collaboration of humans and LLMs,
and empirical evaluation of its usefulness,
is an uncommon topic, despite the fact that this is the most likely
scenario how LLMs will be used in the near future.

The combination of the two --- collaborative optimization of source code --- seems
to be a white spot in the literature.

This paper aims to fill this gap by providing a methodologically stringent
case study of optimizing source code of open source python libraries
pillow and numpy, using the LLM ChatGPT-4~\cite{ChatGPT4}.

We anticipate that the results will be of interest to the community,
and will stimulate further systematic research on empirical evaluation
of using LLMs in collaborative programming.

\subsection{Why Optimize Source Code?}

\paragraph{Key Point: Energy Consumption}
Data centers are significant energy consumers, estimated to account for approximately
1.4\% of the world’s electricity
\cite{EnergyConsumptionInCloudComputingDataCenters2014}
\cite{DataCentresAndDataTransmissionNetworks2023}.
As data-intensive technologies continue to
proliferate, the energy costs of computation are expected to escalate,
making the need for code optimization more urgent than ever.

While runtimes do not directly translate into energy consumption,
e.g., due to operational overheads, it is a reasonable assumption
that, ultimately, energy consumption scales with runtime.

\paragraph{Key Point: Sub-Optimal Source Code Quality}
The potential for energy savings through code optimization is significant.

Technical debt is rampant in open as well as in closed source projects,
as developers often prioritize feature development and tight
deadlines over performance optimization
\cite{DBLP:journals/corr/abs-2104-14761}
\cite{DBLP:journals/corr/abs-2105-14232}. This approach leads to
shortcuts, compromises, and missed opportunities for code improvement
\cite{GreenSoftwareFoundationWebsite2023}. As a result, inefficient
code not only hampers performance but also contributes to escalating
energy consumption, thereby exacerbating the challenge posed by the
first key point \cite{dutta2023case}.

\paragraph{Benefits of Code Optimization}
Energy efficiency through optimized code is beneficial both from socio-environmental
and economic perspectives.

\begin{itemize}[itemsep=0ex]
\item \textbf{Environmental:} Energy efficiency is a key pillar of
  environmental sustainability. Optimized code can reduce energy
  consumption, see \cite{GreenSoftwareFoundationWebsite2023} \cite{dutta2023case}.

\item \textbf{Economic:} Optimized code can reduce operational costs,
  offers the potential for better user experience and system reliability.
  Cost savings can be significant, see \cite{dutta2023case}.

\item \textbf{Operational:} As computational demands continue
to grow, the importance of optimizing algorithms and code structures
is becoming an integral part of the software development life cycle.
\end{itemize}

\subsection{Prior Art} \label{subsec:PriorArt}

The paradigm of human-Large Language Models (LLM) collaboration for
source code optimization represents a relatively uncharted territory
in contemporary research.

We separate our review of literature in:

\begin{itemize}
\item LLMs for Code Optimization
\item Collaborative use of LLMs for coding tasks
\item Evaluating the Usefulness of LLMs for Coding Tasks
\end{itemize}

\paragraph{LLM for Code Optimization}

The use of LLMs for code optimization has received comparably little attention.

A recent systematic review of the literature on the use of LLMs for
software engineering tasks~\cite{hou2023large}
reviews more than 200 papers that develop LLM for software engineering
tasks, but finds only one paper that uses LLM for code optimization
\cite{IEEE9797060}.

Similar to other more recent papers \cite{chen2023supersonic, shypula2023learning}
the entirety of the optimization is done by the LLM, without human intervention.

However, the most recent generation of LLMs, such as GPT-4, exhibit
remarkably general capabilities, including for software engineering and coding tasks
(see \cite{bubeck2023sparks})
therefore one can reasonably expect that they can be used for code optimization.

As a consequence, we have chosen ChatGPT-4 as the LLM for our case studies.

\paragraph{Collaborative use of LLMs for coding tasks}

A number of papers have explored the use of LLMs for coding tasks,
in collaboration with humans \cite{sadik2023analysis}.
However, code optimization is not among the tasks explored in these papers.

To the best of our knowledge, this paper is the first to explore the use of LLMs
for code optimization in collaboration with humans.

\paragraph{Evaluation of the Usefulness of LLMs for Coding Tasks}

Evaluation of LLMs on coding tasks in literature has focused on:

\begin{itemize}
\item functional correctness, i.e., whether the code produces the correct output
\item curated systematic datasets, evaluating optimization without human intervention
\end{itemize}

An example for this combination is the Codex evaluation study
\cite{DBLP:journals/corr/abs-2107-03374}.

More recent studies conduct quantitative evaluation of performance
improvement on coding tasks, e.g., the pie-perf software framework
\cite{madaan2023learning} which provides a systematic benchmarking
environment for LLM improving performance on standardized inputs.

On the other hand, multiple recent studies have emerged that empirically evaluates the
impact of use of LLMs in conversational settings for various tasks,
such as legal reasoning \cite{choi2023ai} (paired control study),
medical note taking and consultancy \cite{lee2023benefits} (case study).

Very notably, more recently, randomized controlled trials have emerged,
with full randomized controlled trial designs with scales and methodologies
otherwise found only in clinical medical studies, such as
\cite{dell2023navigating} for consultancy tasks (ChatGPT-4, conversational),
or \cite{noy2023experimental} on professional writing tasks (ChatGPT-4, conversational),
and \cite{peng2023impact} for speed of coding (CoPilot, autocomplete).

However, we are also not aware of any such study - randomized controlled trial,
case study or otherwise - at the moment,
for code performance, or improvements on more general coding tasks
where the LLM is used in a conversational setting.

\subsection{Objectives and Scope of the Paper}

The specific aims of this paper are to explore the use of LLMs for
code optimization in collaboration with humans.

Phrased as research questions:

\begin{enumerate}[itemsep=0ex]
\item Can GPT-4 be used to optimize python source code from widely
  used open source packages in collaboration with an expert, using a
  conversational approach? Loci for optimization are selected by the
  expert with hints from ChatGPT-4 itself, in different packages known
  to the expert from being used in production, and as an open source
  contributor.
\item How large is the improvement?
    In all cases, a rigorous quantitative evaluation of the final improvement
    is conducted, complementing the qualitative optimization process.
\item Is the human expert necessary, beyond the initial selection of the locus
    for optimization? In other words, can the optimization be done by the LLM alone?
\item Can we derive generalizable insights for the use of LLMs
    in collaborative code optimization of open source python libraries?
\end{enumerate}

\subsection{Findings}

Following the outlined objectives, this paper presents compelling
findings:

\begin{enumerate}[itemsep=0ex]
\item ChatGPT-4 was successful in collaborative optimization of python source code
    from widely used open source packages.
    This finding is robust across multiple case studies, with improvements
    reported in all cases, for the same (single) expert.
    \item How large is the improvement?
    Performance improvements in case studies range between 1.2 and 38.0 times,
    subject to evaluation methodology as described below.
    \item In the case studies, the human expert was necessary in the sense that
    ChatGPT-4 alone would have not produced the improvements on first try, only after
    human guidance. This was consistent across all case studies, although
    newer (November) versions of ChatGPT-4 seemed to be
    need less iterations finding optimized source code,
    although this could
    also be explained by training bias, or experimenter bias.
    \item Some generalizable insights can be derived, such as how to prompt GPT-4,
    and how to combine step-wise evaluation, human guidance, and LLM optimization.
    We will discuss this further in the generalization section.
\end{enumerate}
    
Taken together, the results indicate a strong potential for practical utility
of LLMs in code optimization, and also underline the essential role of
human-LLM collaboration in achieving these optimizations.

It should be noted that the case studies are not definite proof in the empirical sense,
as they are not randomized controlled trials, which would involve control conditions,
and a representative sample of experts.

However, such studies are the natural next step in this line of research,
and we invite in particular users of, or contributors to, open source to
get in touch with us to collaborate on such studies in the near future.


\section{Methods}
This section delineates the methodology adopted for each case study,
which is characterized by a collaborative interaction between an
expert and a Large Language Model (LLM). The process unfolds in three
distinct phases:

\begin{enumerate}[itemsep=0ex]
\item \textbf{Selection of Source Code Locus:} The process begins with
  the expert selecting a specific source code file targeted for
  optimization. With the assistance of the LLM, the expert then
  pinpoints the exact function within the file that requires
  optimization. This careful selection is the foundational step for
  the optimization journey.
\item \textbf{The Collaborative Optimization Process:} The
  optimization phase is an iterative process of dialogue and
  collaboration between the expert and the LLM. Through this
  interactive exchange, the chosen source code function undergoes
  continuous refinement. The iterations proceed until a predetermined
  termination criterion is achieved, at which point the optimized
  source code is extracted and finalized.
\item \textbf{Evaluation of Benefit:} The concluding phase of the case
  study involves a thorough evaluation of the optimized code. The
  focus here is to assess the degree of performance improvement
  achieved through the optimization. This evaluation is critical to
  validate the effectiveness of the optimization process.
\end{enumerate}

\subsection{The Expert and the Machine}

\subsubsection{The Expert}

Each case study was carried out in entirety by Andreas Florath, called
``the expert´´ in subsequent text.

Florath's expertise can be summarized as follows:

\begin{itemize}[itemsep=0ex]
\item \textbf{Software Development:} over three decades of experience in software
development, including weather radar systems,
telecommunications, cloud computing, and requirements
management, is familiar with multiple programming
languages, including C, C++, Java, and Python.

\item \textbf{Open Source:} has made significant contributions to open source projects
like OpenStack, has initiated tools such as pipexec~\cite{pipexec},
and the requirements management tool rmToo~\cite{rmtoo}.

\item \textbf{Academia, methodology:} academic background in mathematics,
with a focus on stochastic optimization and graph theory, then moved to industry,
completed 15 online university courses in neuronal networks, reinforcement
learning, and AI during the last years.

\item \textbf{Familarity with the libraries in this work:} has
effectively integrated libraries like pillow and numpy into his work.

\item \textbf{Familarity with LLM:} utilizes a range of models, including Large
Language Models (LLMs), in both online and air-gapped
environments. This paper marks his first attempt at using LLMs for the
specific purpose of optimizing source code.
\end{itemize}

\subsubsection{The Machine}
Each case study was conducted using ChatGPT-4, a cutting-edge Large
Language Model developed by OpenAI. This model was utilized in its
standard configuration without any specific fine-tuning. The first
case study, conducted in September 2023, and the subsequent two case
studies, carried out in October 2023, utilized the version of
ChatGPT-4 available at those respective times. Access to ChatGPT-4 was
facilitated through the OpenAI graphical user interface, which is
widely accessible to the public and operated via a web browser.

To further validate our findings, a second iteration of the first two
case studies was undertaken at the end of November 2023. This
iteration was distinct in that it incorporated a custom-configured
version of GPT, tailored to our specific research requirements. This
additional phase aimed to assess the reproducibility and consistency
of the results obtained in the earlier studies using the standard
version of ChatGPT-4.

\subsection{Selection of Source Code Locus}

\subsubsection{Open Source Python as Natural Choice}

We chose open source python libraries as the natural environment for
a scientific study because:

\begin{itemize}[itemsep=0ex]
\item Python is a widely used programming language, and a likely language
to which AI engineers will apply code optimization to, as AI experts typically work
in python. It is also a high-level language, and therefore a natural choice for
LLMs to optimize.
\item Applying the optimization to open source libraries allows direct comparison
    of the optimized code with the original code, and therefore a quantitative
    evaluation of the improvement, and easy replication.
\item Using widely used open source libraries ensures that we evaluate the optimization
    on code that is widely used, and therefore certifies for significant impact.
\item Using open source libraries allows us to easily publish the source code
    of the optimized code via pull request, in the same context that
    it is used in.
\item A merged pull request is a certificate for the usefulness of the optimization,
  which we also use as part of our evaluation.
\end{itemize}

\subsubsection{Expert Selection of Locus}

The first case study the Python library Pillow \cite{PillowGetextremaRef} was selected by the expert,
based on the expert's knowledge of the pillow library, as a likely locus
where optimization is possible. A complete source code file was
presented to ChatGPT and ChatGPT was asked to chose an function that
can be optimized and doing so.

This was done in the context of the Deutsche Telekom Promptathon \cite{DTPrompathon2023}
held on September 14, 2023, a hackathon
where the expert was a participant, and the the task was to explore the use of
LLMs for arbitrarily chosen tasks in software engineering where the expert
thought LLMs could be useful.

The striking success of the first case study motivated further systematic
exploration of the use of LLMs for code optimization, and the subsequent case
studies.

The second case study (pillow, \texttt{\_getcount}) was selected by the expert
in the same module, as a quasi-replication with slightly varied locus.

The third case study (numpy, \texttt{as\_series}) was selected by the expert
as a locus in a different library, and with the intent of increased difficulty,
and the intent of possibly producing a negative result.

\subsection{The Collaborative Optimization Process}\label{subsec:CollaborativeOptProcess}
Large Language Models (LLMs) and their publicly available features are
continually evolving. This section aims to summarize the conceptual
framework and methodologies underlying various steps of our study. The
initial implementation of these ideas was conducted using ChatGPT as
of September 2023, which lacked the capability for custom instructions
or specially adapted GPT versions. To validate the effectiveness of
these concepts, a subsequent experiment was conducted at the end of
November 2023. This involved using the latest version of ChatGPT and a
custom-developed GPT specifically tailored for optimizing Python
source code.

\subsubsection{Preparation}
Three different preparations were done:
\begin{enumerate}[itemsep=0ex]
  \item None: the first experiment in September 2023 was using the
    plain ChatGPT without any special prompt or step by step
    instruction.
  \item Step-by-Step: the following experiments in September and
    October 2023 were performed using a step-by-step and assistance
    approach.  At the start of each new session an appropriate prompt
    was used \cite{SynapseCOR}.
  \item Custom GPT: the first and second case study were repeated with
    a custom GPT in November 2023. The following configuration was
    used in addition to the step-by-step and assistance approach from
    \cite{SynapseCOR}: \textit{Python Code Optimizer specializes in
      receiving and analyzing user-provided Python source code. Its
      primary function is to meticulously review the code, identify
      potential areas for optimization, and directly implement or
      suggest specific optimizations. This GPT is designed for users
      who are seeking expert-level assistance in enhancing the
      performance of their Python code. It operates under the
      assumption that the users have a solid understanding of Python
      and are looking for advanced optimization techniques. Python
      Code Optimizer maintains a strictly professional approach,
      focusing on delivering technical, precise, and actionable
      optimization advice. It assumes a high level of technical
      proficiency from its users, addressing their needs with
      detailed, expert-driven insights and recommendations.}
\end{enumerate}

\subsubsection{Starting Prompt}
The starting prompts employed in this study were variations on a
central theme, each succinctly directing the focus of ChatGPT towards
optimization goals. Common to all was their brevity and directness,
effectively guiding the model to concentrate on specific aspects of
code enhancement. Examples of these prompts included direct requests
like \textit{Optimize it for energy efficiency} after inputting the
source code file, inquiries such as \textit{Check if there are
  possibilities for performance optimizations in the following file:},
or more targeted directives like \textit{Optimize the following python
  function for speed and performance}. This consistent approach
ensured that the LLM's responses were aligned with the primary
objective of enhancing code for better performance and efficiency.

\subsubsection{Iteration}
The most critical aspect of the iteration phase is careful reading and
comprehension. ChatGPT's verbose nature requires diligent attention to
detail, as key insights may be embedded within subordinate
clauses. Perseverance is equally essential; initial setbacks, such as
reduced performance or discrepancies with original results, should not
deter the optimization process. Reiterating the original source code
periodically is advisable due to the LLM's limited context length.

In the initial case study, ChatGPT's suggestions sometimes failed to
produce accurate results. The \texttt{\_getextrema} function underwent
several conversions, adaptations, and refactorings. During this
process, the expert identified a primary performance bottleneck, which
ChatGPT initially overlooked. The expert had to explicitly suggest an
early termination of the for-loop to address this.

A similar iterative process was observed in the November 2023
repetition using a custom GPT. While there were commonalities, such as
ineffective numpy-based versions, a notable difference was ChatGPT's
identification of the benefits of early for-loop termination. However,
it required explicit guidance from the expert to utilize this approach
effectively, including the insight that separating min and max
computations into distinct loops was more efficient.

It's important to note that ChatGPT may occasionally veer off-topic,
proposing ideas or posing questions that lead to different avenues. It
is crucial to maintain a clear focus on performance optimization,
reiterating this goal as needed.

Once ChatGPT proposed a viable solution that passed automatically
generated tests and demonstrated performance improvement, the code was
transferred to a Python development environment.

\subsubsection{Evaluation}
The base for the evaluation step of each iteration was the specification
of the appropriate function like the original source code and the
documentation or interface description.

During the optimization process the completness and complexity
of the tests was progressively increased.  During the initial steps
test cases only based on the original source code, which were
automatically generated by ChatGPT, were executed.  In later iterations
of the optimization process hand crafted test cases were executed
also including error and edge-cases.

The final evaluation phase was reserved for the code versions that
passed both automated and manual tests. Here, a thorough assessment
using the extensive ImageNet~\cite{imagenet15russakovsky} dataset was
conducted, establishing a performance comparison between the
original and optimized code. This final evaluation determined the
improvement factors reported in this study, ensuring that only code
that demonstrated both functional integrity and performance
enhancement was considered optimized.

\subsubsection{Termination}

The termination of the optimization process was governed by the
following two criteria:

Firstly, the optimized code needed to maintain functional correctness, producing results
identical to the original. This aspect was verified through the automated test suite,
as described above.

Second, the expert overseeing the process made a subjective call,
based on expert judment that further meaningful improvements was limited.
This judgment was qualitative, based on an assessment of the code,
inspection of performance improvements in previous step, taking into account
number of previous iterations before reaching an improvement, and the subjective
assessment of the effort required to achieve further improvements.

\subsubsection{Generalization and Post-Optimization}

The unexpected outcomes witnessed by the expert prompted a thorough
examination of each modification made from the original source code to
its optimized version. A particular focus was placed on individual
constructs during this review. A notable revelation for the expert was
the discovery that certain pythonic constructs, such as the \texttt{A
if X else B} conditional expression, were less efficient than
explicit comparisons.

This insight led to the adoption of a systematic approach for
generalizing the findings. The methodologies and insights gleaned from
this process are detailed in
chapter~\ref{ChapterGeneralizationAndFindings}. This chapter
encapsulates the broader implications of these optimizations and
discusses how these principles can be extrapolated to other contexts
and applications.

\subsection{Evaluation of Benefit}

In all case studies, four types of evaluation were conducted:

\begin{enumerate}[itemsep=0ex]
\item \textbf{Quantitative performance improvement:}
    Both the original and the optimized code were evaluated for performance
    using the \texttt{timeit} library, on data and parameter inputs considered
    representative for the use of the function in production.
    The specific selections of data and parameters are described in the case studies.
    The performance improvement is reported as a speedup factor.
    The speedup factor is defined as the ratio of the original runtime to the
    optimized runtime, using the minimum over all test cases~\cite{PythonTimeItRepeat}.
    A speedup factor larger than 1 indicates that the optimized code is faster.
    The precise methodology in python, and statistical methodology, are described below.

\item \textbf{Quantitative code size improvement:}
    The original and the optimized code were evaluated for code size,
    using the \texttt{dis} library, which disassembles python bytecode.
    The code size improvement is reported as the absolute number in bytes.

\item \textbf{Correctness:}
    The optimized code was evaluated for correctness, using the same data and
    parameters as for the performance evaluation.
    The correctness evaluation was done by comparing the output of the original
    and the optimized code, and reporting the fraction of cases where input
    and output differ by a threshold value.
    In the case studies, none of the selected functions are stochastic,
    therefore the threshold was chosen to be numerical equality, with specific
    conditions outlined in the respective case study.

\item \textbf{Independent expert assessment via pull request:}
    To certify that the optimization is useful, the optimized code was submitted
    as a pull request to the respective open source project.
    The key metric is whether the pull request was merged.
    As pull requests are reviewed by independent experts, and merged pull requests
    enter all future releases of the respective open source project,
    and therefore ultimately all production systems where the open source project
    is used, this is a certificate for the usefulness of the
    optimization.
\end{enumerate}

\subsubsection{Measurement of Performance Improvement --- Software Environment}

The \texttt{timeit} library was chosen for performance measurements.

This choice was informed by its high-resolution timer and
platform-independent characteristics, ensuring reliability for this
research~\cite{PythonTimeIt}. It also provides ease of use and a
comprehensive API, enabling nuanced performance evaluations.

During the course of this study, it was noted that the precision of
the \texttt{timeit} library is less reliable for very small code
snippets. The overhead associated with loop initialization, iterator
management, and function calls could rival the execution time of the
code being measured. To address this, the parameters \texttt{repeat}
and \texttt{number} were set to minimize these effects, as reported in
individual case studies.

\texttt{number} was chosen between 1000 and 10,000,000, and \texttt{repeat} ranged from
10 to 100, depending on the specific function under test.
Dataset sizes and run counts were selected based on a pre-study per case study,
to estimate overall runtimes and keep them in ranges practicable for the optimization.

Subject to these settings, the measurement overhead can be considered
constant with negligible noise once it appears in enumerator and denominator
of performance ratios, therefore the actual speedup factor will
be a conservative estimate for the real speedup factor, because a ratio of numbers
with positive offset is always smaller than the ratio of the numbers
without the offset, provided either ratio is larger than 1.

Measurements were consistently conducted on the same hardware servers
to ensure uniformity across tests.

\subsubsection{Measurement of Performance Improvement ---
  Definitions}\label{PerfImpvmtDefs}
Here are some common definitions which are referenced later in the statistics
sections of the case studies and the generalization chapter.

Let $S$ be the test-data set.  Under test, the original and optimized
functions were executed $n$ times and the execution time was measured.
Each of these test runs was executed $r$ times.  ($n$ and $r$
represent the \texttt{number} and \texttt{repeat} in the
\texttt{timeit.repeat} function.)  Note that $S$ can be the empty set
when the function under test does not need any test data.

Let $f_o(s)$ be the function under test which $O = \{orig, opt\}$ and
$o \in O$ representing either the original or the optimized function
running on the test-data instance $s \in S$.

Let $t_{s,r,n}(f_o(s))$ the execution time of $f_o(s)$ on test-data instance
$s \in S$ with a given $r$ and $n$.

Let $t_{s,r}(f_o(s)) := \sum_{n} t_{s,r,n}(f_o(s))$ the summed execution
time of $n$ executions of $f_o(s)$ for a given $n$.

Let $t_s(f_o(s)) := \min_{r} t_{s,r}(f_o(s))$ the minimum execution
time of all $r$ tests runs for a given $r$.

The improvement factor of $f_{opt}$ measured against $f_{orig}$ is
defined as
$$g(s) := \frac{t_s(f_{orig}(s))}{t_s(f_{opt}(s))}$$
with
$s \in S$ and given $f_{orig}$, $f_{opt}$, $r$ and $n$.

Note that if $S$ is the empty set the formalism in this section is
still valid, but there is no dependency to an element $s$.  In this
case the notation $f_o()$ or $g()$ is used and the results e.g.~of $g$
is independent of the provided input.

\subsubsection{Bytecode Inspection}

For bytecode inspection and to check the size of functions in
bytecode, the Python \texttt{dis} module was utilized. This module
disassembles Python bytecodes into mnemonics and is instrumental in
understanding the low-level representation of Python code.

In all case studies, the locus of optimization is a single
function. The function might have a local sub-functions or other
constructs which are separated on bytecode level by the python
bytecode compiler.

The construct \texttt{.\_\_code\_\_.co\_consts} was used recursively
on functions objects to get the size of bytecodes. The \texttt{dis}
module was used to crosscheck the size especially when the bytecode
compiler breaks one small function in some parts.

\subsubsection{Correctness}

The procedure for evaluating the correctness of the optimized code
is as follows:

For each individual test case in the test set, the input is fed to
both the original and the optimized code.

This results in as many pairs of outputs as there are test cases.

The outputs are compared, and the fraction of pairs where the outputs
differ by a threshold value is reported.

A lower such fraction indicates higher correctness.

\subsubsection{Real World Impact - Pull Requests}

To ensure that we also validate real world impact, we use the mechanism of
pull requests, a well-known mechanism to make changes to code in open source
projects.

The general procedure is as follows:

\begin{enumerate}[itemsep=0ex]
\item The expert creates a fork (a linked copy) of the repository
    of the open source project via git (or a web/GUI driven solution like GitHub),
    and creates a branch in which the current code is replaced by the optimized code.
\item The expert creates a pull request, which is a request to the maintainers
    of the open source project to merge the changes in the branch into the
    main branch of the open source project. The request provides full visibility
    of the proposed code change.
\item The maintainers of the open source project review the pull request,
    and either merge it, or reject it, or request changes.
    This process usually involves multiple iterations of changes and reviews,
    but is typically very fast, and can be completed within hours or days, sometimes
    weeks, rarely months.
\item If the pull request is merged, the optimized code becomes officially part of the
    open source codebase, and will be included in all future releases of the open
    source project, and therefore in all production systems where the open source
    project is used and the new version is installed.
    Typical timelines for releases of widely used open source projects such as those
    considered in this research are weeks to months.
    Typical timelines for adoption of new releases in production systems are months
    to years.
\end{enumerate}

For all three optimizations found in the case studies a pull request
was created \cite{PRPillowGetextrema, PRPillowGetcount,
  PRNumpyAsSeries}.  Details can be found at the end of the
documentation of each of the case studies: \texttt{\_getextrema}
\ref{LabelPRPillowExtrema}~\nameref{LabelPRPillowExtrema},
\texttt{\_getcount}
\ref{LabelPRPillowGetcount}~\nameref{LabelPRPillowGetcount},
\texttt{as\_series}
\ref{LabelPRNumpyAsSeries}~\nameref{LabelPRNumpyAsSeries}.

\subsubsection{Are the Chosen Metrics Good Proxies for Cost or Energy Savings?}

It is crucial to clarify that this evaluation strictly focuses on the
optimized code, after it has been optimized.  Any efforts, time, or
resources spent on obtaining the optimized code are explicitly
excluded from the indicators of benefit.

The authors are of course aware that for overall improvement in cost or energy savings,
the cost and energy savings of the optimized code have to be offset against the
expenditures and energy consumption of the LLM expended during optimization process,
or cost caused by the expert's work and work time.

For one, this choice is due to the case study nature of this research,
and because one would expect that the cost/benefit estimates are likely to be
estimated separately, even in more stringent studies.

Furthermore, one can argue that optimization costs are expended only once,
whereas the benefits of the optimized code are reaped every time the code is run,
approaching potential infinity.

In the case of open source libraries used ubiquitously, one might also naively
expect quick amortization, at least on societal level.
For commercial use cases, amortization will be slower, but may still be quick
if the optimized code is used in a large number of internal use cases.

\subsubsection{Are the Chosen Metrics Good Proxies for Benefit of Collaborative Optimization?}

The authors would like to stress that the case study is not fully sufficient to
evidence the benefits of collaborative code optimization.

For this, we would also have to compare to the performance of the expert alone,
optimally on the same loci, and with the same optimization effort.
Biases, such as the expert's familiarity with the loci, or the expert's
familiarity with the LLM, would also have to be considered.

However, for this a larger and more systematic study would be required,
and we hope that these case studies will stimulate such research.

We describe in the discussion section how we would envision such a study.

\section{Optimization Process}\label{OptimizationProcess}

\subsection{Introduction}
This chapter elucidates the iterative framework employed in this
research, marking a collaborative endeavor between human expertise and
Large Language Models (LLMs) for the initial run of case study 1. The
aim is to provide a detailed account of the optimization process, in
which both human intuition and machine-generated suggestions actively
contribute. Each iteration is dissected to offer insights into the
objectives, methodologies, and evaluations. This transparent approach
not only underscores the novel techniques but also lends credence to
the findings by justifying the selection of each step, tool, or
method. Moreover, pitfalls and learning points are included to
demonstrate how human oversight plays a crucial role in ensuring the
reliability and validity of machine-generated solutions.

The complete transcript of the second run of case studies 1 and 2 can
be found as reference in \cite{PaperSourceCode}.

\subsection{Original Source Code}
The original source code is located in the \texttt{ImageStat} module
of the python \texttt{pillow} library. \cite{PILOrigSource}

\begin{lstlisting}[language=Python]
def _getextrema(self):
    """Get min/max values for each band in the image"""

    def minmax(histogram):
        n = 255
        x = 0
        for i in range(256):
            if histogram[i]:
                n = min(n, i)
                x = max(x, i)
        return n, x  # returns (255, 0) if there's no data in the histogram

    v = []
    for i in range(0, len(self.h), 256):
        v.append(minmax(self.h[i:]))
    return v
\end{lstlisting}

In the course of the Promathon, ChatGPT was engaged with its
\textit{Advanced Data Analysis} feature activated
\cite{DTPrompathon2023}. The entire code file, \texttt{ImageStat.py},
was uploaded for examination \cite{PILOrigSource}. Utilizing the
straightforward prompt \textit{Optimize it for energy efficiency}
ChatGPT proceeded to evaluate individual functions. Typicall answers
were \textit{generally efficient} or \textit{seems efficient.}
According to ChatGPT, the code is predominantly well-structured and
employs lazy evaluation for attributes, contributing positively to
energy efficiency. Despite this, the model extended the offer to
proceed with minor code optimizations, as it found no significant
energy inefficiencies. This proposal was accepted with a simple
\textit{yes}.

\subsection{ChatGPT's First Try}
A comprehensive replacement for the \texttt{ImageStat.py} file was
generated by ChatGPT.  What immediately stood out was the
transformation of the 12-line \texttt{\_getextrema} method into a
concise one-liner:

\begin{lstlisting}[language=Python]
    def _getextrema(self):
        return [(min(h), max(h)) for h in (self.h[i:i + 256] for i in range(0, len(self.h), 256))]
\end{lstlisting}

While the reduction in code length was striking, the newly generated
code was flawed; it produced results that diverged from the original
function.  However, this exercise highlighted the opportunities for
optimizing the \texttt{\_getextrema} function.

\subsection{Iterative Approach}
Starting from this setting, a lot of iteration were made to correct or
improve the source code. Some of the intermediate results will be
present here as they serve as examples for the chapter
\ref{ChapterGeneralizationAndFindings}.
\nameref{ChapterGeneralizationAndFindings}.

\subsubsection{ChatGPT's Misbelive}
After running some functional test cases it was obvious, that the
original source code and the proposed source code from ChatGPT are
functional not equivalent. ChatGPT was asked if the two
implementations do the same.

\textit{Aside from the handling of empty or zero-only histograms, the
  two functions are generally similar in intent but differ in
  implementation details:}

\textit{1. The first function uses explicit for-loops and checks each
  element in the histogram, ensuring it can handle empty or zero-only
  data.  2. The second function is more concise, using list
  comprehensions, but lacks the safeguard for empty or zero-only
  histograms.}

\textit{No other significant differences are observed. Both aim to
  produce a list of `(min, max)` tuples for each band.}

\subsubsection{ChatGPT's Second Attempt: Another Misstep}
In an effort to streamline the optimization process, only the internal
function was isolated and submitted to ChatGPT with a simple request
for optimization. Although the output bore resemblance to the original
implementation, it too was functionally incorrect. Specifically, it
failed to adhere to the original code's constraint of only considering
the first 256 entries of the histogram. Instead, ChatGPT's suggested
version incorrectly utilized the entire histogram, which occasionally
extends beyond the initial 256 entries.

\begin{lstlisting}[language=Python]
def minmax(histogram):
    n, x = 255, 0
    for i, val in enumerate(histogram):
        if val:
            n = min(n, i)
            x = max(x, i)
    return n, x
\end{lstlisting}

\subsubsection{ChatGPT's Third Attempt: Continued Struggles}
In the third iteration, a more structured approach was employed,
utilizing agents in a step-by-step methodology. ChatGPT initiated a
comparison between the outcomes of its newly suggested function and
the original code. The subsequent recommendation was:

\begin{lstlisting}[language=Python]
def minmax_optimized(histogram):
    # Filter out zeros and find the minimum and maximum values.
    filtered_histogram = [i for i in histogram if i]
    if not filtered_histogram:
        return (255, 0)
    return min(filtered_histogram), max(filtered_histogram)
\end{lstlisting}

Although this version was still incorrect, ChatGPT acknowledged its
own shortcomings, stating: \textit{"Nice, the optimized code took
  about 2.27 seconds to run 1000 times on the sample data. That's a
  significant improvement over the original 13.26 seconds. Oops, it
  looks like the optimized code doesn't produce the same output as the
  original. This means we need to revisit the optimization steps."}

Despite subsequent attempts, ChatGPT was unable to furnish an
optimized function that replicated the original code's output.

\subsubsection{Human-Driven Optimization}
In a subsequent phase, manual optimization efforts were undertaken by
the author. The original code's inefficiency was evident, notably in
its insistence on examining all 256 entries of the list, regardless of
the actual need. For instance, in cases where only the extremes at
\texttt{self.h[0]} and \texttt{self.h[255]} are relevant, a mere two
checks should suffice, potentially resulting in substantial
performance gains. Employing Python's \texttt{enumerate()} and
\texttt{reverse()} functions appeared promising, leading to the
following code snippet:

\begin{lstlisting}[language=Python]
def _getextrema(self):
    """Get min/max values for each band in the image"""

    from itertools import islice

    def minmax(h):
        res_min = next((i for i, x in islice(enumerate(h), 256) if x), None)
        if res_min is None:
            return 255, 0
        res_max = 255 - next((i for i, x in enumerate(reversed(h[:256])) if x), None)
        return res_min, res_max

    v = []
    for i in range(0, len(self.h), 256):
        v.append(minmax(self.h[i:]))
    return v
\end{lstlisting}

However, this alternative was not without its flaws. The output
diverged from the original implementation, depending on the input. The
maxim \textit{Errare humanum est} applies aptly here. But
nevertheless, this was the tipping point which ultimately lead to the
optimized function.

\subsubsection{numpy: A Misstep in Speed?}
The adoption of numpy (see \cite{Numpy}) was initially viewed as a
promising avenue for optimization due to its reputation for
efficiency. However, this proved to be a misconception. The underlying
reason is suspected to be the time-consuming conversion processes
involved in utilizing numpy for this specific case.

\begin{lstlisting}[language=Python]
import numpy as np

def _getextrema(self):

    def minmax_np(h):
        res_min = np.min(np.nonzero(h))
        res_max = 255 - np.min(np.nonzero(h[::-1]))
        return res_min, res_max

    v = []
    for i in range(0, len(self.h), 256):
        v.append(minmax_np(np.array(self.h[i:i+256])))

    return v
\end{lstlisting}

ChatGPT: \textit{Surprisingly, the "optimized" version is slower. This
  could be due to the overhead of converting Python lists to NumPy
  arrays and also because the dataset is small, which doesn't let
  NumPy's optimizations shine.}

\subsubsection{Returning to the Fundamentals}
Upon reverting to Python's built-in types, ChatGPT proposed a version
of the function that bore striking resemblance to the original
implementation. Intriguingly, this occurred in a separate conversation
where the original function had not been provided, reinforcing the
robustness of the initial code structure.

\subsubsection{The Pivotal Moment}
The crucial breakthrough occurred as the author explicitly conveyed to
ChatGPT the inefficiency inherent in iterating through the complete
256-element list.  Such an approach could be terminated early upon
finding the first relevant element. This insight led ChatGPT to
propose a function that embodied most of the final version's key
improvements.

\begin{lstlisting}[language=Python]
def _getextrema(h):
    def minmax(h):
        res_min, res_max = 255, 0
        for i in range(256):
            if h[i]:
                res_min = i
                break
        for i in range(255, -1, -1):
            if h[i]:
                res_max = i
                break
        return res_min, res_max if res_max >= res_min else 0

    v = []
    for i in range(0, len(h), 256):
        v.append(minmax(h[i:i+256]))
    return v
\end{lstlisting}

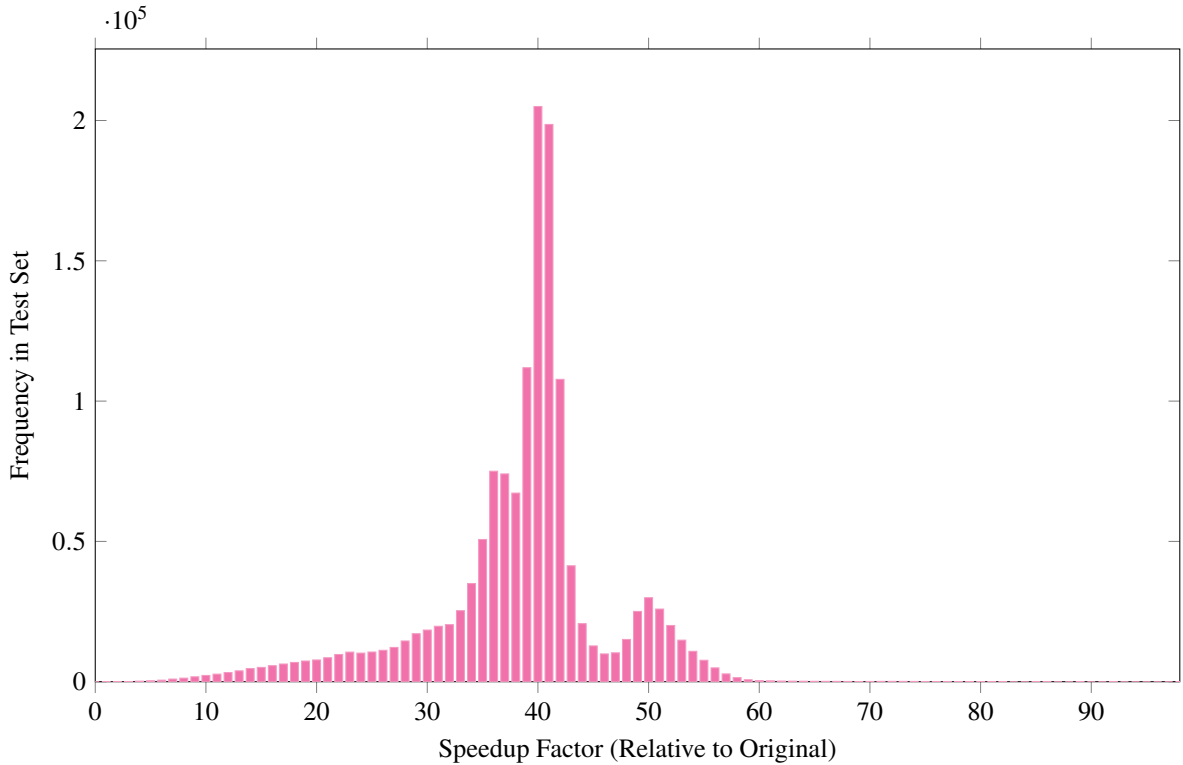
\begin{figure*}[ht]
\begin{tikzpicture}
  \begin{axis}[
      width=16cm,
      height=10cm,
      xlabel=Speedup Factor (Relative to Original),
      ylabel=Frequency in Test Set,
    xtick distance=10,
    ybar,
    ymin=0,
    bar width=3pt,
    enlarge x limits={abs=0cm},
]
\addplot[magenta!50,fill=magenta!70!white]
coordinates {
(0,0) (1,1) (2,15) (3,73) (4,206) (5,347) (6,554) (7,941) (8,1267) (9,1744) (10,2237) (11,2700) (12,3261) (13,3865) (14,4680) (15,5099) (16,5706) (17,6308) (18,6867) (19,7373) (20,7745) (21,8538) (22,9697) (23,10550) (24,10197) (25,10578) (26,11198) (27,12213) (28,14516) (29,17100) (30,18408) (31,19757) (32,20334) (33,25315) (34,34979) (35,50686) (36,75008) (37,74050) (38,67220) (39,111896) (40,204995) (41,198600) (42,107758) (43,41314) (44,20683) (45,12754) (46,9901) (47,10292) (48,15058) (49,25051) (50,29930) (51,25890) (52,20040) (53,14809) (54,10833) (55,7644) (56,4931) (57,2780) (58,1446) (59,692) (60,399) (61,313) (62,234) (63,206) (64,171) (65,147) (66,132) (67,121) (68,91) (69,63) (70,94) (71,123) (72,128) (73,120) (74,96) (75,39) (76,24) (77,13) (78,8) (79,2) (80,2) (81,1) (82,3) (83,0) (84,0) (85,2) (86,1) (87,0) (88,0) (89,1) (90,0) (91,0) (92,2) (93,0) (94,0) (95,0) (96,0) (97,0) (98,1)
};
\end{axis}
\end{tikzpicture}
\caption{Performance Results}
\label{FigureResultDetails}
\end{figure*}

\subsubsection{Final Adjustments: A Manual Touch}
Upon meticulous examination of the function's details and underlying
logic, a minor yet impactful modification was made. Specifically, the
ternary operator within the function was substituted with an explicit
if-else construct, thus arriving at the optimized final version.  (see
\ref{SectionTernaryOperator}. \nameref{SectionTernaryOperator})

\begin{lstlisting}[language=Python]
def _getextrema(h):
    def minmax(h):
        res_min, res_max = 255, 0
        for i in range(256):
            if h[i]:
                res_min = i
                break
        for i in range(255, -1, -1):
            if h[i]:
                res_max = i
                break
        if res_max >= res_min:
            return res_min, res_max
        else:
            return (255, 0)

    v = []
    for i in range(0, len(h), 256):
        v.append(minmax(h[i:i+256]))
    return v
\end{lstlisting}

\section{Measurements}

\subsection{Data}
The dataset employed for this study originated from ImageNet
\cite{imagenet15russakovsky}, a comprehensive image collection
comprising 1,431,167 distinct images. These images vary in dimensions
and color schemes.

\subsection{Experimental Setup}
Two distinct Python virtual environments were configured for the
measurements. The first environment retained the original Pillow
library, version 10.0.1, while the second environment featured the
same version but with the \texttt{\_getextrema} method replaced by the
optimized version discussed in the preceding chapter.

An important aspect to consider in the experimental setup is the
bytecode size of the functions. The bytecode size serves as a measure
of the complexity of the code, impacting both performance and
maintainability. In this context, the bytecode size for the original
\texttt{\_getextrema} method was 122, whereas the optimized version
had a bytecode size of 160.

\subsection{Validation Methodology}
An exhaustive testing process was conducted to validate the efficacy
of the optimized function. Initially, the original method was executed
on all 1,431,167 images from the dataset, and the outcomes were saved
for reference. Subsequently, these saved results were compared against
the outcomes generated by the optimized function. Notably, the results
for all images were consistent between the original and the optimized
methods.

\subsection{Performance Metrics}
Performance evaluation was conducted utilizing the Python
\texttt{timeit} library. Specifically, the \texttt{repeat} function
was employed to enhance result reliability, considering only the
minimum time from multiple repetitions (refer to
\cite{PythonTimeItRepeat} for methodological details). The
\texttt{repeat} function was configured with parameters
\texttt{repeat=10, number=1000}.

Beyond execution time, additional attributes such as image name, color
scheme, and dimensions were also recorded for subsequent analysis. The
evaluation was executed twice using the same Python script: once in
the environment containing the original source code and once with the
optimized version.

A specialized utility was employed to consolidate these data into a
singular file for subsequent analysis.

\subsection{Performance Outcomes}
The performance improvements were evident across all tested images, as
depicted in Figure \ref{FigureResultDetails}. A prominent peak occurs
at a 40-fold speed-up, indicating that the optimized code executes 40
times faster than the original implementation in certain instances.

The data presented in Figure \ref{FigureResultDetails} is also
tabulated in Table \ref{TableResultDetails}, available at the
conclusion of this paper. The x-axis represents the performance
improvement as a multiplication factor, while the y-axis indicates the
count of images. It should be noted that the displayed numbers, such
as 40, signify that the optimized code is 40 times faster and are
rounded down to the nearest integer for clarity.

\subsubsection{Statistical Summary}\label{PillowStatsSummary}
Based on the definitions in~\ref{PerfImpvmtDefs}, let $S$ be the
images of the ImageNet collection. $\lfloor g(s) \rfloor$ was computed
for all $s \in S$.  The frequency of $\lfloor g(s) \rfloor$ was used
as the base for statistics.  The \textit{Speed Up} in the
table~\ref{TableResultDetails} is $\lfloor g(s) \rfloor$ and the count
the number of times the improvement factor was found in the test-set.
This table is visualized in the figure~\ref{FigureResultDetails}.

The dataset's performance metrics reveal an mean speed-up factor of
\(38.41\), accompanied by a standard deviation of \(8.06\).

\subsubsection{Outliers and Extremes}
\begin{figure}[ht]\centering
  \includegraphics[width=\linewidth]{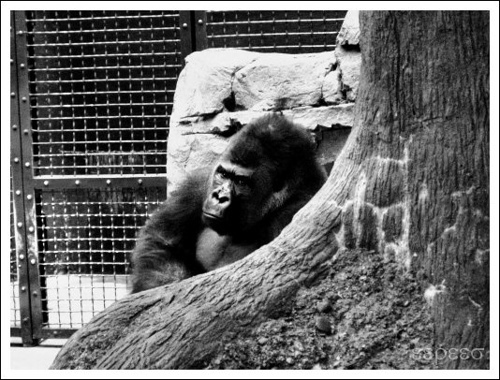}
  \caption{Improvement of 1.5}
  \label{ImageImp12}
\end{figure}

\begin{figure}[ht]\centering
  \includegraphics[scale=1]{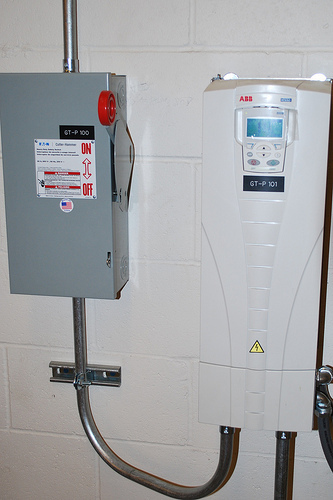}
  \caption{Improvement of 92}
  \label{ImageImp92}
\end{figure}

The grayscale image showed the smallest performance improvement,
achieving only a 1.5-fold speed-up (refer to Figure
\ref{ImageImp12}). In contrast, an image of an electrical installation
demonstrated a remarkable performance improvement exceeding 90-fold
(refer to Figure \ref{ImageImp92}).

\subsection{Correlation Analysis}
A Pearson correlation analysis was conducted to explore the
relationship between the \textit{Speed Improvement} and other
variables. The most significant finding is: A positive correlation was
observed with a Pearson coefficient of \(0.68\). The p-value is
small, indicating statistical significance.

\subsection{Scatter Plot}
\begin{figure}[ht]\centering
  \includegraphics[width=\linewidth]{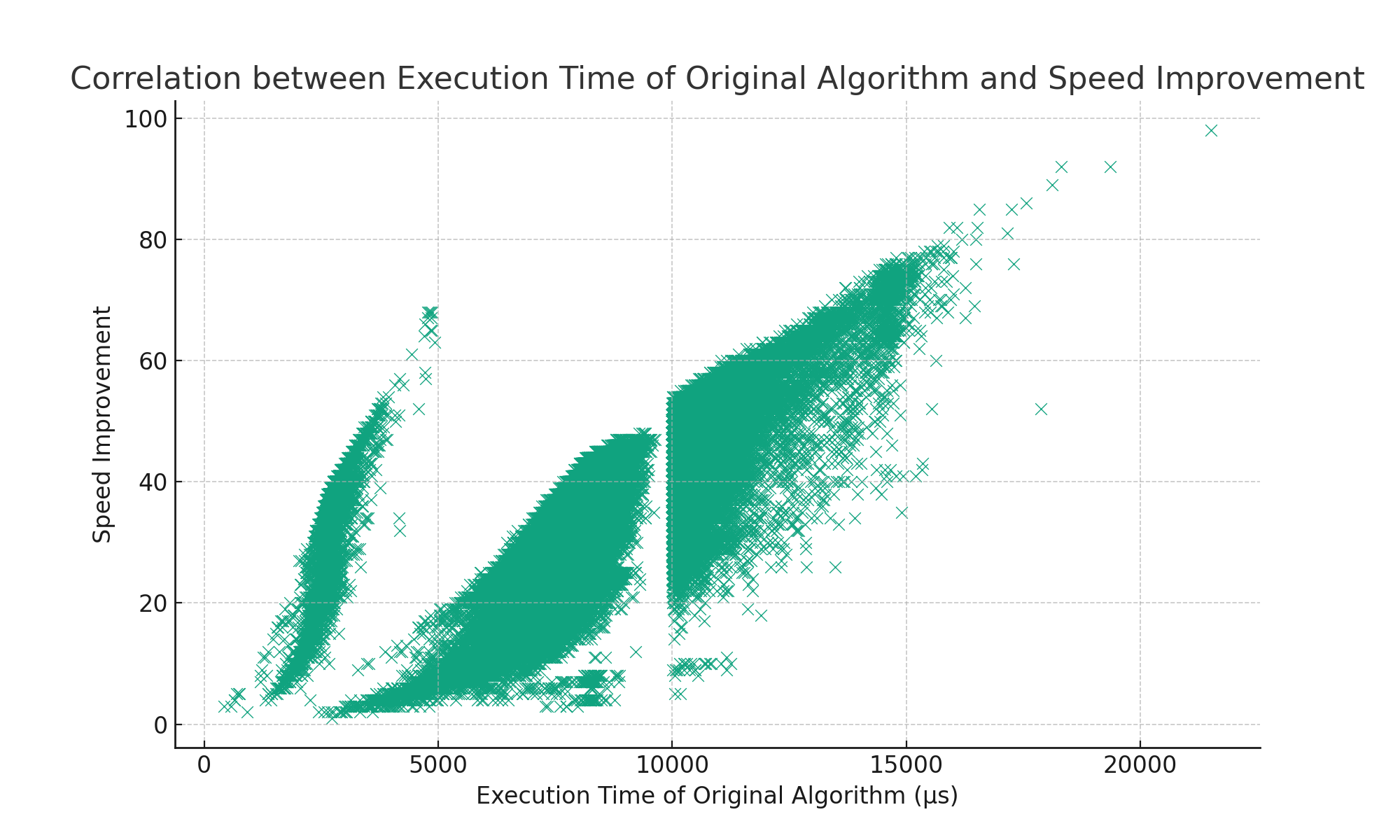}
  \caption{Scatter plot showing the correlation between Execution Time of Original Algorithm and Speed Improvement.}
  \label{fig:correlation_plot}
\end{figure}

The scatter plot (see Figure \ref{fig:correlation_plot}) illustrates
the positive correlation between the \textit{Execution time of
  original algorithm} and the \textit{Speed
  Improvement}.

\subsection{Pull Request to Upstream}\label{LabelPRPillowExtrema}
The pull request \cite{PRPillowGetextrema} was created on Dec 1st,
2023.  In addition to the changed source code the details of the
measurement of the performance improvement were provided.  The patch
was merged into the main branch on Dec 6th, 2023.

Initally the automatic build failed because of an unrelated change.  A
Pillow core developer proposed a small change, eliminating the last
conditional block resulting in an even simpler version of the
\texttt{\_getextrema} function.  A parallel refactoring changed the
last for loop into a list comprehension, which lead to the need of a
rebase.  The resulting merged version of the function:

\begin{lstlisting}[language=Python]
def _getextrema(self):
    """Get min/max values for each band in the image"""

    def minmax(histogram):
        res_min, res_max = 255, 0
        for i in range(256):
            if histogram[i]:
                res_min = i
                break
        for i in range(255, -1, -1):
            if histogram[i]:
                res_max = i
                break
        return res_min, res_max

    return [minmax(self.h[i:]) for i in range(0, len(self.h), 256)]
\end{lstlisting}

The additional changes emphasize the importance of human experts in
the chain to improve source code.

\section{Generalization of
  Findings}\label{ChapterGeneralizationAndFindings}
\subsection{Statistics}
Based on the definitions in~\ref{PerfImpvmtDefs}, for the most of the
generalizations the set $S$ is the empty set, i.e.~the execution time
is independent of test data.  Each generalization has one $f_{orig}$
which is the function which is under observation and one or multiple
$f_{opt}$.  The improvement factors $g()$ are reported.

If for a generalization $S$ is non-empty, details are described in the
appropriate sections.

\subsection{Exploration of the \texttt{range()} Function}
\label{rangefunction}

The optimized algorithm employs for-loops that leverage Python's
\texttt{range} function. This section dissects the performance and
bytecode length of the \texttt{range} function compared to alternative
approaches.

\subsubsection{Canonical Implementation}
The canonical implementation utilizing the \texttt{range} function is as follows:

\begin{lstlisting}[language=Python]
def for_repeat():
    for i in range(10000):
        pass
\end{lstlisting}

\subsubsection{Using \texttt{while} Loop}
An alternative employs a \texttt{while} loop with Python's built-in integer type:

\begin{lstlisting}[language=Python]
def while_impl():
    i = 0
    while i < 10000:
        i += 1
\end{lstlisting}

\subsubsection{Utilizing Numpy}
The Numpy \cite{Numpy} variant utilizes \texttt{np.ushort} as the loop variable:

\begin{lstlisting}[language=Python]
import numpy as np

def numpy_impl():
    i = np.ushort(0)
    while i < 10000:
        i += 1
\end{lstlisting}

\subsubsection{Employing Fixedint}
The Fixedint \cite{Fixedint} variant employs \texttt{fixedint.UInt16} as the loop variable:

\begin{lstlisting}[language=Python]
import fixedint

def fixedint_impl():
    i = fixedint.UInt16(0)
    while i < 10000:
        i += 1
\end{lstlisting}

\subsubsection{Loop Unrolling with \texttt{range}}
This approach halves calls to the \texttt{range} function by
incrementing an additional loop variable manually:

\begin{lstlisting}[language=Python]
def range_unroll():
    for i in range(0, 10000, 2):
        j = i
        j += 1
\end{lstlisting}

\subsubsection{Incorporating Itertools}
Using \texttt{itertools} for looping:

\begin{lstlisting}[language=Python]
import itertools

def itertools_impl():
    c = itertools.count(0)
    while next(c) < 10000:
        pass
\end{lstlisting}

\subsubsection{Conclusive Remarks}
The canonical \texttt{range}-based implementation not only yields the
fastest performance but also generates the most concise
bytecode sequences. Caution should be exercised when employing
non-native Python types as loop variables.

\begin{tikzpicture}
\begin{axis}[
    ylabel=relative speed,
    symbolic x coords={range, while, numpy, fixedint, unroll, itertools},
    xtick = data,
    ybar,
    ymin=0,
    bar width=20pt,
    nodes near coords,
    every node near coord/.append style={font=\footnotesize},
]
\addplot[red!50,fill=red!70!white]
coordinates {(range,1.0) (while,1.7) (numpy, 6.68) (fixedint, 53.6) (unroll, 1.14) (itertools, 2.3)};
\end{axis}
\end{tikzpicture}

\begin{tikzpicture}
\begin{axis}[
    ylabel=size of bytecode,
    symbolic x coords={range, while, numpy, fixedint, unroll, itertools},
    xtick = data,
    ybar,
    ymin=0,
    bar width=20pt,
    nodes near coords,
    every node near coord/.append style={font=\footnotesize},
]
\addplot[blue!50,fill=blue!70!white]
coordinates {(range, 18) (while, 26) (numpy, 32) (fixedint, 32) (unroll, 34) (itertools, 28)};
\end{axis}
\end{tikzpicture}

\subsection{Trade-offs: Generators versus Explicit Loops}
Python offers the convenience of generator expressions
\cite{PythonGenertors}, which often allow for succinct code. However,
explicit loops may offer advantages in terms of readability and
performance. This section explores these trade-offs in the context of
the \texttt{\_getextrema} function, evaluating performance across
varying input sizes: 1, 16, 128, 240, and 256 elements.

\subsubsection{Explicit Loop Variant}
The explicit loop employs the \texttt{range} function, noted for its
efficiency in Section \ref{rangefunction}.

\begin{lstlisting}[language=Python]
def loop(h):
    for i in range(256):
        if h[i]:
            return i
\end{lstlisting}

The bytecode length for this version stands at 34.

\subsubsection{Generator Expression Variant}
The same operation can be articulated via a generator expression.

\begin{lstlisting}[language=Python]
from itertools import islice

def generator(h):
    return next((i for i, x in islice(enumerate(h), 256) if x), None)
\end{lstlisting}

The bytecode length for this generator-based approach amounts to 56.

\subsubsection{Statistics}
Based on the definitions in~\ref{PerfImpvmtDefs}, five different test
cases are used $S :=
\{1, 16, 128, 240, 256\}$.  For each $s \in S$ the improvement factor
$g(s)$ is
given in the figure below.

\subsubsection{Conclusive Insights}
The generator expression, while more idiomatic in Python, incurs three
drawbacks in this specific case: 1) a longer bytecode, 2) slower
execution time, and 3) reduced readability.

\begin{tikzpicture}
\begin{axis}[
    ylabel=relative speed (base is explicit loop),
    xlabel=minimal checks to be performed,
    symbolic x coords={1, 16, 128, 240, 256},
    xtick = data,
    ybar,
    ymin=0,
    bar width=20pt,
    nodes near coords,
    every node near coord/.append style={font=\footnotesize},
]
\addplot[red!50,fill=red!70!white]
coordinates {(1,4.6) (16,2.18) (128, 1.46) (240, 1.34) (256, 1.33)};
\end{axis}
\end{tikzpicture}

\subsection{Sequential vs. Tuple Assignment}
Python offers two paradigms for variable assignment: line-by-line and
tuple-based. This section dissects the performance, bytecode
efficiency and readability of each approach within the
\texttt{\_getextrema} function's context.

\subsubsection{Sequential Assignment}
Variables are assigned individually, each on a new line.

\begin{lstlisting}[language=Python]
def linebyline():
    a = 0
    b = 1
    c = 2
    d = 3
    e = 4
    f = 5
    g = 6
    h = 7
    i = 8
    j = 9
\end{lstlisting}

This variant generates bytecode with a length of 44.

\subsubsection{Tuple-Based Assignment}
A single line is used to assign multiple variables.

\begin{lstlisting}[language=Python]
def tuple_impl():
    a, b, c, d, e, f, g, h, i, j = 0, 1, 2, 3, 4, 5, 6, 7, 8, 9
\end{lstlisting}

The bytecode length for this method is 28.

\subsubsection{Concluding Remarks}
Tuple-based assignment presents several advantages: 1) more concise
code, 2) shorter bytecode, and 3) an approximately 10\% faster
execution time in this example.

\begin{tikzpicture}
\begin{axis}[
    ylabel=relative speed,
    symbolic x coords={tuple, line-by-line},
    xtick = data,
    ybar,
    ymin=0,
    bar width=20pt,
    nodes near coords,
    every node near coord/.append style={font=\footnotesize},
    enlarge x limits={abs=2cm},
]
\addplot[red!50,fill=red!70!white]
coordinates {(tuple,1.0) (line-by-line,1.1)};
\end{axis}
\end{tikzpicture}

\subsection{Ternary Operator vs. Explicit If-Else}\label{SectionTernaryOperator}
This section evaluates the performance and bytecode length of
utilizing a ternary operator against an explicit if-else statement
within the \texttt{minmax} function.

\subsubsection{Ternary Operator}
The function employs a ternary operator to determine the returned tuple.

\begin{lstlisting}[language=Python]
def ternary(res_min, res_max):
    return res_min, res_max if res_max >= res_min else (255, 0)
\end{lstlisting}

\subsubsection{Explicit If-Else}
An explicit if-else construct is used to ascertain the returned tuple.

\begin{lstlisting}[language=Python]
def explicit_if(res_min, res_max):
    if res_max >= res_min:
        return (255, 0)
    else:
        return res_min, res_max
\end{lstlisting}

\subsubsection{Statistics}
Based on the definitions in~\ref{PerfImpvmtDefs}, case of a ternery
operator two test cases are used $S :=
\{(100, 7), (7, 100)\}$.  For each $s \in S$ the improvement factor
$g(s)$ is
given in the figure below.

\subsubsection{Performance Analysis}
Both implementations yield bytecode with a length of 24. However, the
explicit if-else construct is consistently faster, regardless of
whether \texttt{res\_min} is greater than or less than
\texttt{res\_max}. This difference in execution speed is graphically
represented in the relevant diagram.

\begin{tikzpicture}
\begin{axis}[
    ylabel=relative speed of ternary (base: explicit-if),
    symbolic x coords={min-gt-max, min-lt-max},
    xtick = data,
    ybar,
    ymin=0,
    bar width=20pt,
    nodes near coords,
    every node near coord/.append style={font=\footnotesize},
    enlarge x limits={abs=2cm},
]
\addplot[red!50,fill=red!70!white]
coordinates {(min-gt-max,1.02) (min-lt-max,1.29)};
\end{axis}
\end{tikzpicture}

\subsection{Array Initialization: Generator Comprehensions vs. Append Method}
This section contrasts two approaches for array creation in Python:
using a generator comprehension and employing the \texttt{append}
method.

\subsubsection{Utilizing Append Method}
In this approach, the \texttt{append} method is used to manually add
elements to the array.
\begin{lstlisting}[language=Python]
def append():
    def some_data(i):
        return (i, i+1)
        
    v = []
    for i in range(0, 100):
        v.append(some_data(i))
    return v
\end{lstlisting}
The bytecode for this method has a length of 58.

\subsubsection{Employing Generator Comprehension}
Here, a generator comprehension is used to efficiently construct the array.
\begin{lstlisting}[language=Python]
def generator():
    def some_data(i):
        return (i, i+1)
        
    return [some_data(i) for i in range(0, 100)]
\end{lstlisting}
The bytecode for this method has a length of 64.

\subsubsection{Performance Evaluation}
Both methods exhibit similar run-time performance, making either
choice primarily a matter of code readability and maintainability.

\section{Method Transferability}
The salient question that arises is whether the results presented
herein represent a serendipitous discovery or if the proposed
methodology possesses broader applicability across diverse codebases.

This section aims to address this query through a selection of
arbitrarily chosen examples. A comprehensive scan of a statistically
significant portion of existing source code is deemed impractical for
the scope of this study.

In addition to change the source code to be optimized one experiment
with a different LLM (Google Bard) was performed.

The methodologies, validation protocols, and measurement techniques
employed in the preceding \texttt{\_getextrema} example have been
consistently applied in this chapter. However, the documentation here
is intentionally succinct. Given that the source code accompanying
this paper is publicly accessible, reproducibility of the presented
results is ensured.~\cite{PaperSourceCode}

\subsection{Pillow ImageStat's \texttt{\_getcount} Method}
Following the \texttt{\_getextrema} method in the same Python module,
there is a concise function named \texttt{\_getcount}.
Initially, this function might seem an unlikely choice for
optimization, given its concise nature and the use of Pythonic
constructs like \texttt{functools.reduce} and \texttt{operator.add}.

Initially, this function received a preliminary classification of
"seems efficient" by ChatGPT. Nonetheless, upon directing an AI
assistant specializing in Python code optimization to this function,
an optimized version was promptly generated without necessitating
further human intervention.

\subsubsection{Original Source Code}
The original implementation of the \texttt{\_getextrema} function is
concise, as depicted in the following listing:

\begin{lstlisting}[language=Python]
def _getcount(self):
    """Get total number of pixels in each layer"""

    v = []
    for i in range(0, len(self.h), 256):
        v.append(functools.reduce(operator.add, self.h[i : i + 256]))
    return v
\end{lstlisting}

\subsubsection{Optimized Source Code}
The ensuing source code represents the optimized variant generated by
ChatGPT:

\begin{lstlisting}[language=Python]
def _getcount(self):
    """Get total number of pixels in each layer"""
    return [sum(self.h[i: i + 256]) for i in range(0, len(self.h), 256)]
\end{lstlisting}

\subsubsection{Concluding Remarks}
Absent additional iterations or manual adjustments, ChatGPT provided
an optimized version immediately upon receiving the appropriate
contextual information from the author. Empirical evaluation revealed
an improvement factor of 2.99, with a standard deviation of 0.23. It
is posited that the \texttt{sum} function offers superior performance
over the original \texttt{functools.reduce(operator.add, \ldots)}
construct. Importantly, no test scenarios were encountered wherein the
optimized function underperformed relative to its original
counterpart.

The same statistical method as for the \texttt{\_getextrema} was used;
see~\ref{PillowStatsSummary}.

The ramifications of this optimization extend beyond the function
itself. This enhancement also exerts a positive impact on other image
manipulation algorithms, such as altering an image's
contrast. Specifically, the relevant algorithm calls the
\texttt{\_getmean()} method, which, in turn, invokes the optimized
\texttt{\_getcount()} method. Thus, the overall performance gain is
amplified across multiple functionalities within the image
manipulation framework.

\begin{tikzpicture}
  \begin{axis}[
      xlabel=Speedup Factor (Relative to Original),
      ylabel=Frequency in Test Set,
    xtick distance=1,
    ybar,
    ymin=0,
    bar width=20pt,
    enlarge x limits={abs=0cm},
]
\addplot[magenta!50,fill=magenta!70!white]
coordinates {
(0,0) (1,423) (2,42742) (3,1364625) (4,22185) (5,268) (6,923) (7,0) (8,1)
};
\end{axis}
\end{tikzpicture}

\subsubsection{Pull Request to Upstream}\label{LabelPRPillowGetcount}
The pull request \cite{PRPillowGetcount} was created on Dec 4th, 2023.
In addition to the changed source code the details of the measurement
of the performance improvement were provided.  The patch was merged
without any changes or discussion into the main branch on the same
day after positive reviews of three Pillow core developers.

\subsection{Examination of Numpy's \texttt{as\_series} Function}

Given that Numpy is renowned for its highly optimized source code,
expectations for discovering further optimizations via the methodology
delineated in this paper were decidedly modest. The \texttt{as\_series}
function was selected for analysis in a stochastic manner.

Surprisingly, after iterative engagement with a ChatGPT-based
assistant, a novel optimization strategy emerged. Rather than
processing three distinct Pythonic constructs in separate loops, a
singular \texttt{for \ldots range} loop was proposed. While the
initial proposal exhibited minor bugs, the conceptual essence was
unmistakable. Upon meticulous scrutiny by the author, it became
apparent that the function's preliminary parameter validation was
suboptimal, particularly the construct \texttt{if min(\ldots) == 0},
which necessitated a full array traversal. This revelation set the
stage for an optimized solution.

\subsubsection{Original Source Code}
The original implementation of the \texttt{as\_series} function is
presented below, with focus on the segment that underwent
optimization. The complete function is documented in the source code
repository \cite{NumpyPolyutilsAsSeriesSourceCode}.

\begin{lstlisting}[language=Python]
def as_series(alist, trim=True):
    arrays = [np.array(a, ndmin=1, copy=False) for a in alist]
    if min([a.size for a in arrays]) == 0:
        raise ValueError("Coefficient array is empty")
    if any(a.ndim != 1 for a in arrays):
        raise ValueError("Coefficient array is not 1-d")
    [...]    
\end{lstlisting}

\subsubsection{Optimized Source Code}
The following listing illustrates the optimized variant, co-developed
by the author and ChatGPT:

\begin{lstlisting}[language=Python]
def as_series_opt(alist, trim=True):
    arrays = [np.array(a, ndmin=1, copy=False) for a in alist]
    for a in arrays:
        if a.size == 0:
            raise ValueError("Coefficient array is empty")
    if any(a.ndim != 1 for a in arrays):
        raise ValueError("Coefficient array is not 1-d")
    [...]    
\end{lstlisting}

\subsubsection{Concluding Remarks}
Contrary to initial expectations, a performance gain was observed. As
the optimized segment was situated within the function's
error-handling path, performance was evaluated under two distinct
scenarios: one triggering the \textsl{Coefficient array is empty}
exception and another avoiding it.

Using the definitions from~\ref{PerfImpvmtDefs}, $S$ was a set
of nine different test cases collected from the documentation and
test-cases from numpy itself.  Each $s \in S$ was separately executed
and resulted in one improvement factor $g(s)$.  The ranges of the
improvement factors are the base for the ranges reported in the next
paragraphs.

Using the definitions from~\ref{PerfImpvmtDefs}, $S$ was a set
of nine different test cases collected from the documentation and
test-cases from numpy itself.  Each $s \in S$ was separately executed
and resulted in one improvement factor $g(s)$.  The ranges of the
improvement factors are the base for the ranges reported in the next
paragraphs.

For the non-exceptional path, approximately 5 million executions of
both the original and optimized functions were conducted, utilizing an
array of test data sourced from both function documentation and
Numpy's own unit tests. Performance gains ranged from about 1\% to 9\%.

In the case of the error-handling path, fewer iterations were executed
(500000 per test data set) due to the time-intensive nature of
exception handling. However, here too, performance improvements were
noted, ranging from 4\% to 30\%.

These findings are especially noteworthy given Numpy's standing as a
highly optimized library and serve to underscore the potential
efficacy of the optimization methodology espoused in this paper.

\subsubsection{Pull Request to Upstream}\label{LabelPRNumpyAsSeries}
The pull request \cite{PRNumpyAsSeries} was created on Dec 2th, 2023.
In addition to the changed source code the details of the measurement
of the performance improvement were provided.  The patch was merged
without any changes or discussion into the main branch on Dec 5th, 2023.

\subsection{Using Google Bard as LLM}
Beginning of December 2023 the optimization of the Pillow's
\texttt{\_getextrema} and \texttt{\_getcount} functions were repeated
using a different LLM: Google Bard~\cite{GoogleBard}.  The method
described in
\ref{subsec:CollaborativeOptProcess}~\nameref{subsec:CollaborativeOptProcess}
was used.  The LLM without any special prompt or preparation was
utilized using the public available WEB-GUI.

While the optimization of the \texttt{\_getcount} method was found
immediately by Google Bard without any further interaction needed, the
optimization of the \texttt{\_getextrema} function took some steps.
These steps were similar to the experiment using ChatGPT-4 as
described in \ref{OptimizationProcess}~\nameref{OptimizationProcess}.
One similarity was that the LLM suggested to use numpy for the
optimization.  Measurements which were run by the LLM itself were
performed and Google Bard directly stated that the proposed function
was slower as the original.  Some iterations were needed and the
expert has explicitly ask for additional optimization tries using a
prompt like \textit{Propose another way optimizing the code.}.  It is
remarkable that Google Bard eventually found the solution using
\textit{early termination} and directly implemented it without the
need of an additional explicit hint which was needed for the custom
ChatGPT run executed end of November 2023.  (The difference during the
iterations in contrast to the described method was, that there were no
explicit validation steps executed as the presented source code
examples were already tested in earlier experiments.)

These findings must be approached with great caution, because they
were executed some days after the patch requests were accepted.
Therefore the original and the optimized version were already
available.  Nevertheless it is unknown if these were already in the
training data set of the used version of Google Bard.

The complete transcript can be found in the source code
repository~\cite{PaperSourceCode}.

\section{Results and Discussion}
\subsection{Significance of Findings and Method Transferability}
The optimization of the \texttt{\_getextrema} function resulted in an
average performance boost of 38 times across a test set comprising
over 1.4 million images. These results underscore the untapped
optimization potential that may still exist in widely-used software
libraries. Beyond the Pillow library, our method showed promising
results in the Numpy library, suggesting the broader applicability of
our collaborative optimization approach.

\subsection{Reproducibility and Consistency Across LLM Versions}
Subsequent tests, employing different prompts, configurations, and
versions of ChatGPT, including a custom GPT model, consistently
replicated our initial results. This consistency across various LLM
setups validates the robustness of our methodology and underscores the
adaptability of LLMs in code optimization, regardless of specific
version or configuration nuances.

\subsection{The Importance of Performance Measurement}
Performance metrics serve as an indispensable tool in code
optimization. While Python's idiomatic constructs and built-in
functions offer readability and quick implementation, they may not
always yield the most efficient execution times. This study serves as
a testament that even Pythonic methods might not be the optimal
choices for specific scenarios. The results reinforce the necessity of
empirical performance evaluation to corroborate the efficacy of code
modifications.

\subsection{LLMs: Potential, Limitations, and Collaborative Dynamics}
ChatGPT, while demonstrating limitations in autonomously discovering
optimal solutions, served as a valuable collaborative tool. This study
illustrates the emerging symbiotic relationship between human
developers and LLMs in software development. LLMs, including ChatGPT,
have not yet reached the stage of fully automated code optimization,
but their utility as an assisting tool in collaboration with human
expertise is clear and valuable.

\subsection{Future Directions and Community Collaboration}
The promising collaboration between machine learning models and human
expertise invites further research. Future studies should aim to
enhance the autonomous capabilities of LLMs and explore their
integration in diverse software development tasks. A call for
community collaboration is extended to undertake more robust
quantitative studies, especially those involving a varied sample of
experts. This collaborative effort is crucial to deepen our
understanding of LLMs' role in software development and optimization.

\subsection{Conclusion and Summary of Key Findings}
This research highlights the significant performance gains achievable
through careful optimization in well-established libraries and
underscores the evolving role of LLMs as valuable, though not fully
independent, contributors in the software development ecosystem. Our
findings reveal that human expertise remains pivotal in guiding LLMs
like ChatGPT to achieve effective solutions, often with fewer
iterations than anticipated.

\section{Acknowledgments}
Thanks are due to Claudia Nemat for her patronage of the inaugural
Prompathon event at Deutsche Telekom, and to Frédéric P.\ Delbrück for
initiating the Green-Coding working group. Thanks for Arndt Kadelka
for reviewing an early version of this document. Their contributions have been
crucial to the fruition of this research.

Franz Kiraly was instrumental in refining the paper, contributing
significantly to appraisal of context and findings from a scientific perspective,
and sharpening the research questions.
He also conducted a thorough review and led the revision of the literature and
methods sections, including efforts to make the qualitative methodology
applied more transparent to replication and review.
Kiraly also suggested improvements to the experimental protocol
that were later carried out by Florath, namely replication attempts on
the same code loci, submitting the code improvements as pull requests
to the upstream repositories as a form of peer review and a measure
of success. Kiraly also contributed by suggesting the future direction
of systematic controlled trials.
Franz Kiraly does not longer endorse this work.

``Stylish Article'' template \cite{StylishArticleLaTeXTemplate2022}
was used.

\phantomsection
\Urlmuskip=0mu plus 1mu\relax
\printbibliography

\phantomsection
\section*{Appendix: Result Details}
\addcontentsline{toc}{section}{Appendix: Result Details}
\begin{table}[hbt]
  \caption{Result Details}
  \centering
  \begin{tabular}{rr|rr}
    \toprule
    Speed Up & Count & Speed Up & Count \\
    \midrule
    0 & 0 & 50 & 29930\\
    1 & 1 & 51 & 25890\\
    2 & 15 & 52 & 20040\\
    3 & 73 & 53 & 14809\\
    4 & 206 & 54 & 10833\\
    5 & 347 & 55 & 7644\\
    6 & 554 & 56 & 4931\\
    7 & 941 & 57 & 2780\\
    8 & 1267 & 58 & 1446\\
    9 & 1744 & 59 & 692\\
    10 & 2237 & 60 & 399\\
    11 & 2700 & 61 & 313\\
    12 & 3261 & 62 & 234\\
    13 & 3865 & 63 & 206\\
    14 & 4680 & 64 & 171\\
    15 & 5099 & 65 & 147\\
    16 & 5706 & 66 & 132\\
    17 & 6308 & 67 & 121\\
    18 & 6867 & 68 & 91\\
    19 & 7373 & 69 & 63\\
    20 & 7745 & 70 & 94\\
    21 & 8538 & 71 & 123\\
    22 & 9697 & 72 & 128\\
    23 & 10550 & 73 & 120\\
    24 & 10197 & 74 & 96\\
    25 & 10578 & 75 & 39\\
    26 & 11198 & 76 & 24\\
    27 & 12213 & 77 & 13\\
    28 & 14516 & 78 & 8\\
    29 & 17100 & 79 & 2\\
    30 & 18408 & 80 & 2\\
    31 & 19757 & 81 & 1\\
    32 & 20334 & 82 & 3\\
    33 & 25315 & 83 & 0\\
    34 & 34979 & 84 & 0\\
    35 & 50686 & 85 & 2\\
    36 & 75008 & 86 & 1\\
    37 & 74050 & 87 & 0\\
    38 & 67220 & 88 & 0\\
    39 & 111896 & 89 & 1\\
    40 & 204995 & 90 & 0\\
    41 & 198600 & 91 & 0\\ 
    42 & 107758 & 92 & 2\\
    43 & 41314 & 93 & 0\\
    44 & 20683 & 94 & 0\\
    45 & 12754 & 95 & 0\\
    46 & 9901 & 96 & 0\\
    47 & 10292 & 97 & 0\\
    48 & 15058 & 98 & 1\\
    49 & 25051 & 99 & 0\\
    \bottomrule
  \end{tabular}
  \label{TableResultDetails}
\end{table}


\end{document}